%% file: manuscript.tex
\documentclass[preprint,12pt]{elsarticle}
\usepackage[T1]{fontenc}
\usepackage{amsmath,amssymb,bm}
\usepackage{booktabs,tabularx,multirow}
\usepackage{graphicx}
\usepackage{placeins}
\usepackage{siunitx}
\usepackage{xcolor}
\usepackage{microtype}
\usepackage{hyperref}
\usepackage[nameinlink,capitalise]{cleveref}

\journal{Computer Physics Communications}

\hypersetup{
  colorlinks=true,
  linkcolor=blue!55!black,
  citecolor=blue!55!black,
  urlcolor=blue!55!black
}
\newcommand{\AccelNet}{\textsc{AccelNet}}
\newcommand{\aenet}{{\ae}net}
\newcommand{\nTwoPTwo}{\texttt{n2p2}}
\newcommand{\dd}{\mathrm{d}}

\newcommand{\vect}[1]{\bm{#1}}

\begin{document}

\begin{frontmatter}

\title{AccelNet: Exact backward-compatible acceleration of polynomial angular descriptors through Cartesian moment factorization}

\author[aff1,aff2]{Yuki Nagai\corref{cor1}}
\ead{nagai.yuki@mail.u-tokyo.ac.jp}
\cortext[cor1]{Corresponding author}
\affiliation[aff1]{organization={Information Technology Center, The University of Tokyo},
                   addressline={6--2--3 Kashiwanoha},
                   city={Kashiwa},
                   state={Chiba},
                   postcode={277--0882},
                   country={Japan}}
\affiliation[aff2]{organization={Department of Advanced Materials Science, The University of Tokyo},
                   city={Kashiwa},
                   state={Chiba},
                   postcode={277--8561},
                   country={Japan}}

\begin{abstract}
We present \AccelNet, an exact, backward-compatible method for accelerating
existing trained \aenet\ and \nTwoPTwo\ neural-network potentials without
retraining.  For angular terms with separable one-neighbor weights and a
finite polynomial dependence on \(\cos\theta\), the method exploits their
hidden finite-rank structure to replace explicit neighbor-pair loops by
one-neighbor Cartesian moments.  \AccelNet\ reads
models trained with either package and reproduces their descriptors, energies,
and analytic forces to floating-point roundoff.  We verified this equivalence
for H$_2$O and TiO$_2$ models and tested the resulting potentials in LAMMPS
molecular-dynamics simulations.  \AccelNet\ also enables GPU-accelerated
molecular dynamics with existing \aenet\ and \nTwoPTwo\ potentials.
Moment evaluation achieves speedups of up
to 10.9 on a CPU and 15.8 on a GPU relative to direct evaluation within
\AccelNet.  The implementation, model-conversion tools,
and LAMMPS interfaces are released as open-source software.
\end{abstract}


\end{frontmatter}

\section{Introduction}
\label{sec:introduction}

Molecular dynamics (MD) enables atomistic simulations of a wide range of
systems, including inorganic and organic materials as well as biological
matter, by numerically integrating the equations of motion of their
constituent atoms \cite{rahman1964correlations}.  Such simulations require a
model of the potential-energy surface from which the forces acting on the
atoms can be obtained.
Traditionally, these forces have been described using empirical interatomic
potentials designed for particular classes of materials and interactions
\cite{stillinger1985silicon,cornell1995amber}.
Advances in electronic-structure methods and computational resources have
made first-principles MD increasingly accessible, in which the energy is
evaluated quantum mechanically and the atomic forces are obtained from its
derivatives \cite{car1985unified}.  This approach incorporates electronic effects into the
interatomic interactions and thereby enables more predictive atomistic
simulations than are generally possible with fixed empirical functional
forms.

Machine-learning interatomic potentials provide a route to retaining much of
the accuracy of first-principles calculations at substantially lower
computational cost.  Instead of solving the electronic-structure problem at
every MD step, these models learn an approximation to the reference
potential-energy surface from energies and forces calculated for a set of
representative atomic configurations.  After training, energies and forces
can be evaluated rapidly while closely reproducing the underlying
first-principles data \cite{behler2007generalized,bartok2010gap,zhang2018deep}.
More recently, foundation models trained on large and chemically diverse
first-principles data sets have extended this approach across broad ranges of
elements, compositions, and atomic environments.  Models such as MACE-MP-0
and MatterSim have demonstrated that a single pretrained potential can
approach the accuracy of its density-functional-theory reference data for a
wide variety of materials \cite{batatia2025foundation,yang2024mattersim}.
These models also provide a starting point for system-specific fine-tuning
\cite{nagai2026target}.

The broad chemical coverage of foundation models is achieved by increasing
both the diversity of the training data and the capacity of the model.  For
production MD, however, accuracy and transferability are not the only
considerations: the cost of each force evaluation directly limits the system
sizes and trajectory lengths that can be reached.  Compact, system-specific
potentials therefore remain complementary to large general-purpose models,
because a comparatively small network can provide the throughput required for
large-scale and long-time simulations.  An important example is the
high-dimensional neural-network potential introduced by Behler and Parrinello,
which represents the total energy as a sum of environment-dependent atomic
energies \cite{behler2007generalized,behler2011symmetry}.  Software packages
such as \textsc{RuNNer}~2.0, \aenet, and \nTwoPTwo\ provide optimized
implementations of this atom-centered neural-network strategy and interfaces
to molecular-dynamics codes
\cite{knoll2026runner,artrith2016aenet,singraber2019lammps,singraber2019n2p2,chen2021aenet}.
Such models consequently remain important when simulation throughput is the
primary requirement and an accurate potential has already been constructed
for the system of interest.

Despite their large speed advantage over first-principles MD, conventional
machine-learning potentials remain substantially more expensive to evaluate
than many highly optimized empirical potentials.  Comparative studies of
neural-network, Gaussian-process, spectral, and moment-based potentials have
demonstrated a pronounced trade-off between prediction accuracy and
computational cost \cite{zuo2020assessment,xie2023ultrafast}.  For compact
atom-centered neural-network potentials, one important source of this cost is
the construction of angular descriptors.  The Chebyshev expansion of the
atom-centered angular distribution function used by \aenet\ and the angular
symmetry functions available in \nTwoPTwo\ depend on a central atom \(i\) and
two of its neighbors, \(j\) and \(k\)
\cite{behler2011symmetry,artrith2017chebyshev,chen2021aenet}.  At first sight,
this pairwise dependence appears unavoidable because each angular contribution
depends simultaneously on two neighbors through the bond angle
\(\theta_{ijk}\).  Direct evaluation therefore enumerates neighbor pairs: an
environment containing \(z_i\) neighbors requires \(z_i(z_i-1)/2\) angular-pair
evaluations.  For
\(z_i=100\), this amounts to 4950 distinct neighbor pairs for a single central
atom, with contributions evaluated for each applicable angular component.
The resulting quadratic dependence on the local neighbor count can dominate
the evaluation even when the neural network itself is small.

In this work, we focus on the mathematical structure of the angular
descriptors used by \aenet\ and \nTwoPTwo\ and derive an equivalent
representation that removes the explicit neighbor-pair loop for the Chebyshev
angular descriptors in \aenet\ and a subset of the angular symmetry functions
in \nTwoPTwo, specified in \cref{sec:n2p2-polynomial-angular}
\cite{behler2011symmetry,artrith2017chebyshev,singraber2019n2p2}.  The descriptor
values obtained with the new representation are mathematically identical to
the original definitions; the numerical agreement is assessed in
\cref{sec:verification}.  We further developed \AccelNet, an inference engine
that reads trained \aenet\ and \nTwoPTwo\ models, evaluates their descriptors
using the new representation, and computes the corresponding energies and
analytic forces.  Because the descriptor definitions, scaling, network
architecture, and trained weights are preserved, existing models can be used
without retraining and accelerated without changing the accuracy of the
original potential.  \AccelNet\ is also implemented as a LAMMPS potential,
enabling the accelerated evaluation to be used directly in production MD
simulations \cite{plimpton1995lammps}.  We evaluate the method on CPUs and
GPUs and compare the GPU performance of LAMMPS coupled to \AccelNet\ with
CabanaMD \cite{desai2022cabanamd} for a published aqueous-interface model.

\section{Neural-network potentials and hidden finite-rank descriptor evaluation}
\label{sec:theory}

\subsection{Atomic-energy decomposition and local descriptors}
\label{sec:atomic-energy-model}

In an atom-centered neural-network potential, the geometry surrounding each
atom is first mapped to a fixed-length descriptor vector.  Let
\(\mathcal{X}=\{(\vect{r}_i,\alpha_i)\}_{i=1}^{N}\) denote an atomic configuration,
where \(\vect{r}_i\) and \(\alpha_i\) are the position and chemical species of atom
\(i\), respectively.  Its local descriptor is written as
\begin{equation}
  \vect{G}_i(\mathcal{X})
  = \left(\vect{G}_i^{\mathrm{rad}},
           \vect{G}_i^{\mathrm{ang}}\right),
  \label{eq:descriptor-vector}
\end{equation}
where the radial components describe the distribution of neighbor distances
and the angular components describe correlations between pairs of bonds in the
local environment.  This division into radial and angular information is
common to the original Behler--Parrinello construction, \aenet, and \nTwoPTwo,
although the functional forms and channel conventions differ among them
\cite{behler2007generalized,behler2011symmetry,artrith2016aenet,
artrith2017chebyshev,singraber2019n2p2}.

The descriptor of atom \(i\) is supplied to a neural network associated with
its central-atom species.  The network returns an atomic contribution
\begin{equation}
  E_i = \mathcal{F}_{\alpha_i}\!\left(\vect{G}_i;\vect{\theta}_{\alpha_i}\right),
  \label{eq:atomic-energy-network}
\end{equation}
where \(\vect{\theta}_{\alpha_i}\) denotes the weights and biases of the network for
species \(\alpha_i\).  The total potential energy is the sum of these atomic
contributions,
\begin{equation}
  E_{\mathrm{NN}}(\mathcal{X};\vect{\theta})
  = \sum_{i=1}^{N}
    \mathcal{F}_{\alpha_i}\!\left(\vect{G}_i(\mathcal{X});
                                  \vect{\theta}_{\alpha_i}\right)
  = \sum_{i=1}^{N} E_i .
  \label{eq:total-energy-decomposition}
\end{equation}
This atomic-energy decomposition is the defining construction of the
high-dimensional neural-network potential introduced by Behler and Parrinello
\cite{behler2007generalized}.  The individual \(E_i\) are latent model
quantities rather than separately assigned DFT targets; it is their sum that is
fitted to the reference total energy.

For a training set of configurations \(\mathcal{X}_m\) with DFT reference
energies \(E_m^{\mathrm{DFT}}\), the network parameters are determined by
minimizing an energy loss, schematically,
\begin{equation}
  \vect{\theta}^{\star}
  = \underset{\vect{\theta}}{\operatorname{arg\,min}}
    \sum_m w_m
    \left[
      E_{\mathrm{NN}}(\mathcal{X}_m;\vect{\theta})
      - E_m^{\mathrm{DFT}}
    \right]^2,
  \label{eq:energy-training-loss}
\end{equation}
where \(w_m\) denotes any configuration-dependent weighting.  Practical
training procedures may additionally include forces and stresses in the loss.
Once trained, atomic forces are obtained by differentiating the summed energy,
\begin{equation}
  \vect{F}_a
  = -\frac{\partial E_{\mathrm{NN}}}{\partial \vect{r}_a}.
  \label{eq:nn-force-definition}
\end{equation}

\subsection{General form of radial and angular descriptors}
\label{sec:radial-angular-descriptors}

Both radial and angular descriptors characterize the neighborhood of a central
atom \(i\) within a finite cutoff radius.  For a cutoff
\(R_{\mathrm{c}}\), define
\begin{equation}
  \mathcal{N}_i(R_{\mathrm{c}})
  = \left\{j\ne i\,\middle|\,r_{ij}<R_{\mathrm{c}}\right\},
  \label{eq:cutoff-neighborhood}
\end{equation}
where
\begin{equation}
  \vect{r}_{ij}=\vect{r}_j-\vect{r}_i,
  \qquad
  r_{ij}=\lVert\vect{r}_{ij}\rVert,
  \label{eq:aenet-bond-vector}
\end{equation}
and let \(\alpha_i\) denote the chemical species of atom \(i\).  A typical
radial descriptor component can be written in the general form
\begin{equation}
  G_{ip}^{\mathrm{rad}}
  = \sum_{j\in\mathcal{N}_i(R_{\mathrm{c}}^{\mathrm{rad}})}
    \Phi_p^{\mathrm{rad}}
    \!\left(r_{ij};\alpha_i,\alpha_j\right).
  \label{eq:general-radial-descriptor}
\end{equation}
It is a sum of contributions from individual neighbors, and each contribution
usually depends only on the central and neighboring species and on their
separation \(r_{ij}\).

An angular descriptor contains information about two neighbors \(j\) and
\(k\) of the same central atom.  A general three-atom contribution may depend
on the three distances and the bond angle,
\begin{align}
  G_{iq}^{\mathrm{ang}}
  ={}& \sum_{\substack{j<k\\
                       j,k\in\mathcal{N}_i(R_{\mathrm{c}}^{\mathrm{ang}})}}
    \Phi_q^{\mathrm{ang}}
    \!\left(r_{ij},r_{ik},r_{jk},c_{ijk};
            \alpha_i,\alpha_j,\alpha_k\right),
  \label{eq:general-angular-descriptor}\\
  c_{ijk}={}&\cos\theta_{ijk}
  =\frac{\vect{r}_{ij}\cdot\vect{r}_{ik}}{r_{ij}r_{ik}}.
  \label{eq:general-angle-cosine}
\end{align}
Specific descriptor families need not use every argument in
\cref{eq:general-angular-descriptor}; for example, some omit \(r_{jk}\).
The essential distinction from \cref{eq:general-radial-descriptor} is that an
angular contribution is associated with a pair of neighbors and therefore
requires a sum over \((j,k)\).

As a concrete example, consider the Chebyshev descriptor introduced by
Artrith, Urban, and Ceder and implemented in \aenet\
\cite{artrith2016aenet,artrith2017chebyshev}.  Its radial component of
Chebyshev order \(n\) is
\begin{equation}
  G_{in\gamma}^{\mathrm{rad}}
  = \sum_{\substack{j\ne i\\r_{ij}<R_{\mathrm{c}}^{\mathrm{rad}}}}
    w_{\alpha_j}^{(\gamma)}
    f_{\mathrm{c}}\!\left(r_{ij};R_{\mathrm{c}}^{\mathrm{rad}}\right)
    T_n\!\left(\frac{2r_{ij}}{R_{\mathrm{c}}^{\mathrm{rad}}}-1\right),
  \label{eq:aenet-radial-chebyshev}
\end{equation}
where \(T_n\) is the Chebyshev polynomial of the first kind and
\(w_{\alpha_j}^{(\gamma)}\) is the weight assigned to the species of neighbor
\(j\) in channel \(\gamma\).  The radial coordinate is mapped from
\([0,R_{\mathrm{c}}^{\mathrm{rad}}]\) to the Chebyshev interval \([-1,1]\).
The factor \(f_{\mathrm{c}}\) smoothly restricts the local environment and is
implemented in \aenet\ as the cosine cutoff
\begin{equation}
  f_{\mathrm{c}}(r;R_{\mathrm{c}})=
  \begin{cases}
    \dfrac{1}{2}\left[
      \cos\!\left(\dfrac{\pi r}{R_{\mathrm{c}}}\right)+1
    \right], & r<R_{\mathrm{c}},\\[6pt]
    0, & r\ge R_{\mathrm{c}}.
  \end{cases}
  \label{eq:aenet-cosine-cutoff}
\end{equation}
The structural channel has
\(w_{\alpha}^{(\mathrm{str})}=1\), whereas the element-sensitive channel uses the
species-dependent weights defined by the descriptor setup.  Equation
\eqref{eq:aenet-radial-chebyshev} contains one sum over the neighbors of atom
\(i\).

Here we follow the angular convention in the \aenet\ implementation used
for comparison: the Chebyshev argument is \(\cos\theta_{ijk}\).
The corresponding angular component is
\begin{align}
  G_{in\gamma}^{\mathrm{ang}}
  ={}& \sum_{\substack{j<k\\
                       r_{ij}<R_{\mathrm{c}}^{\mathrm{ang}}\\
                       r_{ik}<R_{\mathrm{c}}^{\mathrm{ang}}}}
    w_{\alpha_j}^{(\gamma)}w_{\alpha_k}^{(\gamma)}
    f_{\mathrm{c}}\!\left(r_{ij};R_{\mathrm{c}}^{\mathrm{ang}}\right)
    f_{\mathrm{c}}\!\left(r_{ik};R_{\mathrm{c}}^{\mathrm{ang}}\right)
    T_n(c_{ijk}).
  \label{eq:aenet-angular-chebyshev}
\end{align}
Thus, unlike the radial descriptor, evaluation of one angular component
requires an explicit loop over all unordered neighbor pairs \((j,k)\).  If
\(z_i\) atoms lie within the angular cutoff of atom \(i\), the number of pair
contributions is
\begin{equation}
  N_{\mathrm{pair}}(i)=\binom{z_i}{2}
  =\frac{z_i(z_i-1)}{2}.
  \label{eq:aenet-neighbor-pair-count}
\end{equation}
The radial accumulation therefore grows linearly with \(z_i\), whereas the
direct angular accumulation grows quadratically and repeats the pair loop for
the required polynomial orders and channels.

\subsection{Exact moment factorization of polynomial angular descriptors}
\label{sec:polynomial-moment-factorization}

We now consider an angular descriptor whose dependence on the bond angle is a
finite polynomial of \(c_{ijk}=\cos\theta_{ijk}\).  We further assume that the
descriptor has no explicit dependence on \(r_{jk}\), and that its remaining
radial and chemical factors separate into one-neighbor weights.  For a channel
\(\gamma\), the descriptor then has the form
\begin{equation}
  G_{i\gamma}^{\mathrm{ang}}
  = \sum_{\substack{j<k\\j,k\in\mathcal{N}_i}}
    h_{ij}^{(\gamma)}h_{ik}^{(\gamma)}P(c_{ijk}),
  \label{eq:polynomial-angular-assumption}
\end{equation}
where \(h_{ij}^{(\gamma)}\) contains the cutoff, radial, and species-dependent
factors associated with neighbor \(j\), and
\begin{equation}
  P(c)=\sum_{p=0}^{P_{\max}} a_p c^p
  \label{eq:angular-power-polynomial}
\end{equation}
is a polynomial of finite degree \(P_{\max}\).  The Chebyshev angular
descriptor in \cref{eq:aenet-angular-chebyshev} satisfies these assumptions
because \(T_n(c)\) is a polynomial of degree \(n\)
\cite{artrith2017chebyshev}.

To expose the separability of each power of \(c_{ijk}\), write the bond
directions as unit vectors,
\begin{equation}
  \vect{u}_{ij}=\frac{\vect{r}_{ij}}{r_{ij}},
  \qquad
  c_{ijk}=\vect{u}_{ij}\cdot\vect{u}_{ik}.
  \label{eq:unit-bond-vector-moment}
\end{equation}
The Cartesian expansion of this dot product makes the dependence on neighbors
\(j\) and \(k\) explicit.  For a nonnegative integer \(p\), the multinomial
theorem gives
\begin{align}
  c_{ijk}^{p}
  ={}&
  \left(
    u_{ij,x}u_{ik,x}
    +u_{ij,y}u_{ik,y}
    +u_{ij,z}u_{ik,z}
  \right)^p \nonumber\\
  ={}&
  \sum_{\substack{\ell_x,\ell_y,\ell_z\ge0\\
                   \ell_x+\ell_y+\ell_z=p}}
  \frac{p!}{\ell_x!\ell_y!\ell_z!}
  \left(
    u_{ij,x}^{\ell_x}
    u_{ij,y}^{\ell_y}
    u_{ij,z}^{\ell_z}
  \right)
  \left(
    u_{ik,x}^{\ell_x}
    u_{ik,y}^{\ell_y}
    u_{ik,z}^{\ell_z}
  \right).
  \label{eq:dot-power-multinomial}
\end{align}
Each term is a product of a function of neighbor \(j\) and the corresponding
function of neighbor \(k\).  This finite separation permits the sum over
neighbors to be performed before the products are formed.

For the multi-index
\(\boldsymbol{\ell}=(\ell_x,\ell_y,\ell_z)\), define the Cartesian moment
\begin{equation}
  M_{i,\boldsymbol{\ell}}^{(\gamma)}
  = \sum_{j\in\mathcal{N}_i}
    h_{ij}^{(\gamma)}
    u_{ij,x}^{\ell_x}
    u_{ij,y}^{\ell_y}
    u_{ij,z}^{\ell_z}.
  \label{eq:cartesian-neighbor-moment}
\end{equation}
Using \cref{eq:dot-power-multinomial}, the sum over all ordered neighbor pairs
is
\begin{align}
  B_{ip}^{(\gamma)}
  &\equiv
  \sum_{j,k\in\mathcal{N}_i}
    h_{ij}^{(\gamma)}h_{ik}^{(\gamma)}c_{ijk}^{p}
  \nonumber\\
  &=
  \sum_{\substack{\ell_x,\ell_y,\ell_z\ge0\\
                   \ell_x+\ell_y+\ell_z=p}}
  \frac{p!}{\ell_x!\ell_y!\ell_z!}
  \left(M_{i,\boldsymbol{\ell}}^{(\gamma)}\right)^2.
  \label{eq:ordered-pair-moment-contraction}
\end{align}
This ordered sum contains the diagonal terms \(j=k\), which are absent from
the original descriptor.  Since
\(c_{ijj}=\vect{u}_{ij}\cdot\vect{u}_{ij}=1\), their contribution is
\begin{equation}
  D_i^{(\gamma)}
  = \sum_{j\in\mathcal{N}_i}
    \left(h_{ij}^{(\gamma)}\right)^2.
  \label{eq:moment-self-term}
\end{equation}
The two off-diagonal ordered terms \((j,k)\) and \((k,j)\) are equal.
Consequently,
\begin{equation}
  \sum_{\substack{j<k\\j,k\in\mathcal{N}_i}}
  h_{ij}^{(\gamma)}h_{ik}^{(\gamma)}c_{ijk}^{p}
  =\frac{1}{2}\left[B_{ip}^{(\gamma)}-D_i^{(\gamma)}\right].
  \label{eq:power-pair-sum-moment}
\end{equation}
Substitution into \cref{eq:polynomial-angular-assumption} yields the exact
moment representation
\begin{equation}
  G_{i\gamma}^{\mathrm{ang}}
  =\frac{1}{2}\left[
    \sum_{p=0}^{P_{\max}} a_p
    \sum_{\substack{\ell_x,\ell_y,\ell_z\ge0\\
                     \ell_x+\ell_y+\ell_z=p}}
    \frac{p!}{\ell_x!\ell_y!\ell_z!}
    \left(M_{i,\boldsymbol{\ell}}^{(\gamma)}\right)^2
    -P(1)D_i^{(\gamma)}
  \right].
  \label{eq:exact-polynomial-moment-descriptor}
\end{equation}
No approximation has been introduced: the pair sum and the moment expression
contain exactly the same terms.  For fixed polynomial degree, the moments in
\cref{eq:cartesian-neighbor-moment} are accumulated in a single loop over the
\(z_i\) neighbors.  Their number depends on \(P_{\max}\), but not on \(z_i\),
so the dependence on the neighbor count is reduced from the quadratic pair
enumeration to a linear moment accumulation.

\subsection{Chebyshev angular descriptors}
\label{sec:chebyshev}

For the \aenet\ Chebyshev angular descriptor in
\cref{eq:aenet-angular-chebyshev}, define
\begin{equation}
  h_{ij}^{(\gamma)}
  =w_{\alpha_j}^{(\gamma)}
   f_{\mathrm{c}}\!\left(r_{ij};R_{\mathrm{c}}^{\mathrm{ang}}\right).
  \label{eq:chebyshev-one-neighbor-weight}
\end{equation}
The angular component can then be written as
\begin{equation}
  G_{in\gamma}^{\mathrm{Ch}}
  =\sum_{\substack{j<k\\j,k\in\mathcal{N}_i}}
   h_{ij}^{(\gamma)}h_{ik}^{(\gamma)}T_n(c_{ijk}).
  \label{eq:chebyshev-angular}
\end{equation}
The Chebyshev polynomial \(T_n\) has degree \(n\), and hence has a finite
power-basis expansion
\begin{equation}
  T_n(c)=\sum_{p=0}^{n}t_{np}c^p.
  \label{eq:chebyshev-power-expansion}
\end{equation}
The coefficients can be generated directly from
\(T_{n+1}(c)=2cT_n(c)-T_{n-1}(c)\):
\begin{align}
  t_{0p}&=\delta_{p0},\nonumber\\
  t_{1p}&=\delta_{p1},\nonumber\\
  t_{n+1,p}&=2t_{n,p-1}-t_{n-1,p},
  \label{eq:chebyshev-coefficient-recurrence}
\end{align}
where coefficients with indices outside \(0\le p\le n\) are taken to be
zero.

Applying \cref{eq:exact-polynomial-moment-descriptor} with
\(a_p=t_{np}\) gives
\begin{equation}
  G_{in\gamma}^{\mathrm{Ch}}
  =\frac{1}{2}\left[
    \sum_{p=0}^{n}t_{np}
    \sum_{\substack{\ell_x,\ell_y,\ell_z\ge0\\
                     \ell_x+\ell_y+\ell_z=p}}
    \frac{p!}{\ell_x!\ell_y!\ell_z!}
    \left(M_{i,\boldsymbol{\ell}}^{(\gamma)}\right)^2
    -D_i^{(\gamma)}
  \right],
  \label{eq:auc-moment-form}
\end{equation}
where the self term simplifies because \(T_n(1)=1\).  This expression is
algebraically identical to \cref{eq:chebyshev-angular}; it changes only the
order in which the neighbor contributions are accumulated.

\subsection{Applicable angular symmetry functions in \nTwoPTwo}
\label{sec:n2p2-polynomial-angular}

Among the atom-centered symmetry functions available in \nTwoPTwo, the
present factorization applies to the wide angular exponential symmetry
function \cite{behler2011symmetry,singraber2019n2p2}.  For a central atom
\(i\) and a prescribed pair of neighbor species \((\beta,\delta)\), let
\(\mathcal{P}_i^{\beta\delta}\) denote the unordered neighbor pairs
\((j,k)\), with \(j<k\), whose species are \(\beta\) and \(\delta\) in either
order.  The corresponding descriptor is
\begin{align}
  G_{i,\beta\delta}^{\mathrm{wide}}
  ={}&2^{1-\zeta}
  \sum_{(j,k)\in\mathcal{P}_i^{\beta\delta}}
  \left(1+\lambda c_{ijk}\right)^{\zeta}
  \exp\!\left\{-\eta\left[
    (r_{ij}-r_{\mathrm{s}})^2+(r_{ik}-r_{\mathrm{s}})^2
  \right]\right\}
  \nonumber\\[-2pt]
  &\hspace{34mm}\times
  f_{\mathrm{c}}(r_{ij};R_{\mathrm{c}})
  f_{\mathrm{c}}(r_{ik};R_{\mathrm{c}}).
  \label{eq:n2p2-wide-angular}
\end{align}
Unlike the narrow angular symmetry function, \cref{eq:n2p2-wide-angular}
contains no factor depending on the neighbor--neighbor distance \(r_{jk}\).
All factors other than the bond-angle dependence can therefore be separated
into one-neighbor weights.  Define
\begin{equation}
  h_{ij}^{(\beta)}
  =\chi_{\beta}(\alpha_j)
   \exp\!\left[-\eta(r_{ij}-r_{\mathrm{s}})^2\right]
   f_{\mathrm{c}}(r_{ij};R_{\mathrm{c}}),
  \label{eq:n2p2-one-neighbor-weight}
\end{equation}
where \(\chi_{\beta}(\alpha_j)\) is one when neighbor \(j\) has species
\(\beta\), and zero otherwise.  Equation
\eqref{eq:n2p2-wide-angular} is equivalently
\begin{equation}
  G_{i,\beta\delta}^{\mathrm{wide}}
  =\frac{2^{1-\zeta}}{1+\delta_{\beta\delta}}
   \sum_{\substack{j,k\in\mathcal{N}_i\\j\ne k}}
   h_{ij}^{(\beta)}h_{ik}^{(\delta)}
   \left(1+\lambda c_{ijk}\right)^{\zeta}.
  \label{eq:n2p2-wide-angular-ordered}
\end{equation}
The denominator accounts for the two equivalent orderings when
\(\beta=\delta\); for distinct species, the two species masks select only one
ordering of each physical pair.

If \(\zeta\) is a positive integer, the angular factor is the finite
polynomial
\begin{equation}
  \left(1+\lambda c\right)^{\zeta}
  =\sum_{p=0}^{\zeta}
   \binom{\zeta}{p}\lambda^p c^p.
  \label{eq:n2p2-angular-binomial}
\end{equation}
The same Cartesian separation used for the Chebyshev descriptor can then be
applied term by term.  Define the species-resolved moments
\begin{equation}
  M_{i,\boldsymbol{\ell}}^{(\beta)}
  =\sum_{j\in\mathcal{N}_i}
   h_{ij}^{(\beta)}
   u_{ij,x}^{\ell_x}u_{ij,y}^{\ell_y}u_{ij,z}^{\ell_z}
  \label{eq:n2p2-species-moment}
\end{equation}
and their contraction at degree \(p\),
\begin{equation}
  B_{ip}^{(\beta\delta)}
  =\sum_{\substack{\ell_x,\ell_y,\ell_z\ge0\\
                    \ell_x+\ell_y+\ell_z=p}}
   \frac{p!}{\ell_x!\ell_y!\ell_z!}
   M_{i,\boldsymbol{\ell}}^{(\beta)}
   M_{i,\boldsymbol{\ell}}^{(\delta)}.
  \label{eq:n2p2-mixed-moment-contraction}
\end{equation}
The diagonal contribution included in this contraction is
\begin{equation}
  D_i^{(\beta\delta)}
  =\sum_{j\in\mathcal{N}_i}
   h_{ij}^{(\beta)}h_{ij}^{(\delta)}.
  \label{eq:n2p2-mixed-self-term}
\end{equation}
Substituting \cref{eq:n2p2-angular-binomial} into
\cref{eq:n2p2-wide-angular-ordered} therefore gives
\begin{equation}
  G_{i,\beta\delta}^{\mathrm{wide}}
  =\frac{2^{1-\zeta}}{1+\delta_{\beta\delta}}
   \left[
    \sum_{p=0}^{\zeta}
     \binom{\zeta}{p}\lambda^p
     B_{ip}^{(\beta\delta)}
    -(1+\lambda)^{\zeta}D_i^{(\beta\delta)}
   \right].
  \label{eq:n2p2-wide-angular-moment}
\end{equation}
For \(\beta\ne\delta\), the species masks are disjoint and
\(D_i^{(\beta\delta)}=0\).  For \(\beta=\delta\), the self term removes the
\(j=k\) contributions and the denominator converts the ordered sum into the
original unordered pair sum.  Thus, as in the Chebyshev case, the original
\nTwoPTwo\ descriptor is recovered exactly from moments accumulated in a
single loop over the neighbors.  Angular functions containing an
\(r_{jk}\)-dependent exponential or cutoff, and functions with noninteger
\(\zeta\), do not satisfy the assumptions of \cref{sec:polynomial-moment-factorization}
and remain on the direct evaluation path.

\subsection{Relation to moment-based and atomic-density expansions}
\label{sec:relation-to-ace}

The factorization derived above is closely related to established moment-based
and atomic-density representations.  Moment tensor potentials (MTPs) first
accumulate Cartesian tensors of the schematic form
\begin{equation}
  \mathcal{M}_{\mu\nu}(i)
  =\sum_{j\in\mathcal{N}_i}
   f_{\mu\nu}(r_{ij},\alpha_i,\alpha_j)
   \vect{r}_{ij}^{\otimes\nu},
  \label{eq:mtp-moment-tensor}
\end{equation}
and construct rotational invariants by contracting products of these tensors
\cite{shapeev2016mtp}.  The moments in
\cref{eq:cartesian-neighbor-moment} use normalized bond directions and an
explicit Cartesian monomial representation, but the underlying operation---a
one-neighbor sum followed by tensor contraction---is the same.

The atomic cluster expansion (ACE) begins from species-resolved neighbor
densities
\begin{equation}
  \rho_i^{(\beta)}(\vect{r})
  =\sum_{j\in\mathcal{N}_i}
   \chi_{\beta}(\alpha_j)
   \delta\!\left(\vect{r}-\vect{r}_{ij}\right)
  \label{eq:ace-neighbor-density}
\end{equation}
and its projections onto one-particle basis functions,
\begin{equation}
  A_{iv}^{(\beta)}=\langle\phi_v,\rho_i^{(\beta)}\rangle
  =\sum_{j\in\mathcal{N}_i}
   \chi_{\beta}(\alpha_j)\phi_v(\vect{r}_{ij}).
  \label{eq:ace-density-projection}
\end{equation}
Products of these projections are then coupled to the required rotational
symmetry.  This ``density trick'' replaces explicit sums over neighbor tuples
by products of quantities that have already been summed over individual
neighbors, giving linear scaling with neighbor count at fixed basis size
\cite{drautz2019ace,dusson2022ace}.  CACE expresses the same general strategy
through a Cartesian atomic-density expansion, making its relation to
\cref{eq:cartesian-neighbor-moment,eq:dot-power-multinomial} particularly close
\cite{cheng2024cace}.  In the present three-body setting, the subtraction of
\cref{eq:moment-self-term} explicitly removes the self contributions introduced
when products of one-neighbor sums are formed.

These connections mean that neither Cartesian moments nor the finite feature
representation of a polynomial dot-product kernel is claimed here as a new
general principle.  MTP, ACE, and CACE ordinarily define a basis or model that
is subsequently fitted.  The objective here is different: moments are used
only as an internal evaluation of descriptor components already fixed by a
trained legacy model.  The required identity is
\begin{equation}
  \vect{G}_i^{\mathrm{moment}}
  =\vect{G}_i^{\mathrm{legacy}},
  \label{eq:drop-in-equality}
\end{equation}
including the original pair convention, self-term removal, species channels,
descriptor ordering, scaling, and network weights.  Thus, the contribution of
the present work is to identify and exploit the hidden finite-rank structure
of established polynomial angular descriptors as an exact, backward-compatible
evaluation algorithm, rather than to introduce another descriptor family.

\subsection{Computational cost}
\label{sec:computational-cost}

The cost of an angular descriptor is controlled primarily by two local
quantities: the number \(z_i\) of neighbors of atom \(i\) inside the angular
cutoff and the maximum polynomial degree \(P_{\max}\) required by the
descriptor.  In the direct evaluation, all unordered neighbor pairs are
visited.  Their number is
\begin{equation}
  N_{\mathrm{pair}}(i)=\binom{z_i}{2}
  =\frac{z_i(z_i-1)}{2}.
  \label{eq:cost-number-of-pairs}
\end{equation}
If the powers or polynomial orders through \(P_{\max}\) are generated by a
recurrence for each pair, the leading work per angular channel is therefore
\begin{equation}
  W_{\mathrm{direct}}(i)
  =O\!\left[N_{\mathrm{pair}}(i)P_{\max}\right]
  =O\!\left(z_i^2P_{\max}\right).
  \label{eq:direct-computational-cost}
\end{equation}
The prefactor depends on the number of angular descriptors and on how many
radial and chemical factors can be shared, but the quadratic dependence on
\(z_i\) remains.

In the moment evaluation, the neighbor-pair loop is replaced by the
accumulation of Cartesian monomials.  At a fixed degree \(p\), the number of
nonnegative triples \((\ell_x,\ell_y,\ell_z)\) satisfying
\(\ell_x+\ell_y+\ell_z=p\) is
\begin{equation}
  N_p=\binom{p+2}{2}.
  \label{eq:number-of-moments-at-degree}
\end{equation}
Consequently, the number of moments required through degree \(P_{\max}\) is
\begin{equation}
  N_{\mathrm{mom}}(P_{\max})
  =\sum_{p=0}^{P_{\max}}\binom{p+2}{2}
  =\binom{P_{\max}+3}{3}.
  \label{eq:number-of-moments-through-degree}
\end{equation}
Accumulating these moments for all neighbors costs
\(O[z_iN_{\mathrm{mom}}(P_{\max})]\).  Their contractions and the final
linear combinations of polynomial coefficients do not introduce another
factor of \(z_i\).  The leading cost is thus
\begin{equation}
  W_{\mathrm{moment}}(i)
  =O\!\left[(z_i+1)\binom{P_{\max}+3}{3}\right]
  =O\!\left(z_iP_{\max}^3\right)
  \label{eq:moment-computational-cost}
\end{equation}
for large \(z_i\) and \(P_{\max}\).  The memory required for the accumulated
moments has the corresponding
\(O[N_{\mathrm{mom}}(P_{\max})]\) dependence.

These estimates show why neither method is uniformly faster.  Increasing the
cutoff radius or the local atomic density increases \(z_i\), and therefore
penalizes the direct pair loop quadratically but the moment accumulation only
linearly.  By contrast, raising the polynomial degree increases the direct
work approximately linearly, whereas the number of three-dimensional
Cartesian moments grows as \(P_{\max}^3/6\).  Ignoring implementation-dependent
constants, the ratio of the two leading costs scales approximately as
\begin{equation}
  \frac{W_{\mathrm{direct}}}{W_{\mathrm{moment}}}
  =O\!\left(\frac{z_i}{P_{\max}^2}\right).
  \label{eq:cost-ratio}
\end{equation}
Moment evaluation is therefore favored by large neighborhoods and moderate
polynomial degree, while direct evaluation can remain preferable for small
\(z_i\) or high degree.  In an actual model, the crossover also depends on the
number of distinct radial, cutoff, species, and chemical-weight groups:
moments must be accumulated separately for each such group, but can be reused
by all descriptors that share it.  The same distinction governs analytical
force evaluation.  Direct differentiation traverses the neighbor pairs,
whereas differentiating the moment form retains a one-neighbor accumulation
at fixed polynomial degree.

\section{Numerical verification of the exact reformulation}
\label{sec:verification}

The derivation in \cref{sec:polynomial-moment-factorization} establishes equality in
exact arithmetic.  A practical implementation must additionally preserve all
descriptor conventions, chemical channels, cutoff factors, scaling
transformations, neural-network parameters, and analytical derivatives.  We
therefore verified the reformulation at three levels.  First, direct and
moment evaluation were compared for individual descriptor values and their
Cartesian derivatives.  Second, complete energies and forces were compared
with the original \aenet\ and \nTwoPTwo\ implementations using the same trained
models and atomic configurations.  Third, analytical forces from the complete
moment path were checked by finite differences of the predicted energy.  No
model was retrained or refitted for these comparisons.

\subsection{Test models and systems}

Three models were selected to cover both descriptor families and substantially
different angular environments.  The H$_2$O Chebyshev model provides a
short-cutoff, low-neighbor-count case.  The TiO$_2$ Chebyshev model has the
same maximum angular order but a longer cutoff and a much larger number of
neighbor pairs.  A second TiO$_2$ model contains the applicable \nTwoPTwo\
angular symmetry functions introduced in \cref{sec:n2p2-polynomial-angular}.
The TiO$_2$ Chebyshev benchmark uses anatase, whereas the \nTwoPTwo\
benchmark uses a rutile configuration with atomic displacements from the
ideal lattice sites.  The model dimensions and configurations used for the numerical comparisons are
summarized in \cref{tab:models}.  GPU measurements of the \nTwoPTwo\ TiO$_2$
model use 192, 5184, and 24,000 atoms, obtained by replicating the same
24-atom cell.  Its 54 angular functions per element share three radial
parameter groups and use powers \(\zeta=1,2,4\).

We assess the GPU performance of \AccelNet's direct evaluator using the
MoS$_2$-confined water potential of Schran et al.\ \cite{schran2021aqueous}.
For this benchmark, we compare LAMMPS coupled to \AccelNet\ with CabanaMD.
The selected network is member
\texttt{nnp-001} of the published model set.  Each of its four elements
(Mo, S, O, and H) has 80 radial and angular symmetry functions and two
20-node hidden layers.  The angular functions contain the neighbor--neighbor
distance and are evaluated by direct pair summation.  The starting cell
contains 56 MoS$_2$ units and 109 water molecules, giving 495 atoms;
\(2\times2\times2\) and \(3\times3\times3\) replicas contain 3960 and
13,365 atoms.  The published descriptor parameters and network weights are
used throughout.

\begin{table}[t]
  \centering
  \small
  \setlength{\tabcolsep}{3.5pt}
  \caption{Benchmark models and representative system sizes.  Network widths
  include the input and scalar output layers.  Larger replicas are used in
  the CPU/GPU comparisons.}
  \label{tab:models}
  \begin{tabularx}{\linewidth}{@{}>{\raggedright\arraybackslash}p{0.17\linewidth}rrr>{\raggedright\arraybackslash}p{0.18\linewidth}X@{}}
    \toprule
    System & \(N\) & Desc. & \(r_c^{\mathrm{ang}}\) (\AA) & Network & Details \\
    \midrule
    H$_2$O Chebyshev & 1152 & 52 & 3.175 & 52--25--25--1 & radial order 20; angular order 4 \\
    TiO$_2$ Chebyshev & 1152 & 44 & 6.5 & 44--15--15--1 & radial order 16; angular order 4 \\
    TiO$_2$ \nTwoPTwo & 192 & 70 & 6.5 & 70--15--15--1 & 16 radial (type 2); 54 polynomial angular (type 9) \\
    MoS$_2$/H$_2$O & 495 & 80 & 6.35 & 80--20--20--1 & radial G2 and angular G4 \\
    \bottomrule
  \end{tabularx}
\end{table}

\subsection{Energy and force equivalence}

For a calculation path \(X\), we quantify the differences from the relevant
external reference by
\begin{align}
  \Delta E_X &= \left|E_X-E_{\mathrm{ref}}\right|,\\
  \Delta F_X &= \max_{a\mu}
  \left|F_{a\mu}^{X}-F_{a\mu}^{\mathrm{ref}}\right|,
  \label{eq:validation-errors}
\end{align}
where \(a\) labels atoms and \(\mu\in\{x,y,z\}\).  The reference is \aenet\
for the Chebyshev models and \nTwoPTwo\ for the second-generation TiO$_2$
model.  Both the direct and moment paths in \AccelNet\ use the same loaded
network, descriptor scaling, and atomic reference energies.  Consequently,
their comparison with the external program tests the complete route from
model input to energy and force output, while the comparison between the two
\AccelNet\ paths isolates the moment reformulation.

\begin{table}[t]
  \centering
  \small
  \setlength{\tabcolsep}{4pt}
  \setlength{\tabcolsep}{3.5pt}
  \caption{Maximum initial-frame absolute differences against the external
  reference over five trials per evaluation method in the single-core
  measurements.  Energy differences refer to
  the total energy in eV; force differences are in eV/\AA.  Comparisons use
  identical atomic coordinates.  Zeros indicate identical energies in
  outputs recorded to 17 significant digits.}
  \label{tab:validation}
  \begin{tabular}{@{}llllrr@{}}
    \toprule
    Suite & Reference & Compared path & Atoms & \(\max|\Delta E|\) & \(\max|\Delta F|\) \\
    \midrule
    H$_2$O Chebyshev & \aenet & direct & 1152 & 0 & \(3.35\times10^{-14}\) \\
                     & \aenet & moment & 1152 & 0 & \(3.38\times10^{-14}\) \\
    TiO$_2$ Chebyshev & \aenet & direct & 1152 & 0 & \(9.49\times10^{-15}\) \\
                      & \aenet & moment & 1152 & 0 & \(1.07\times10^{-14}\) \\
    TiO$_2$ \nTwoPTwo\ model & \nTwoPTwo & direct & 192 & \(2.91\times10^{-11}\) & \(1.62\times10^{-13}\) \\
                   & \nTwoPTwo & moment & 192 & \(2.91\times10^{-11}\) & \(2.64\times10^{-13}\) \\
    \bottomrule
  \end{tabular}
\end{table}

For the Chebyshev models, the maximum force difference is
\(3.38\times10^{-14}\)~eV/\AA.  For the \nTwoPTwo\ TiO$_2$ model,
the total-energy difference is \(2.91\times10^{-11}\)~eV and the maximum
force difference is \(2.64\times10^{-13}\)~eV/\AA.

\subsection{Descriptor derivatives and finite-difference tests}

\input{verification-details}

\subsection{CPU and GPU energy and force verification}
\label{sec:gpu-verification}

The TiO$_2$ production calculations were checked against standard
\nTwoPTwo\ at identical atomic coordinates, using full-precision output.
Each measured process was checked at its initial configuration; the final
configuration from the first trajectory of each method and size was also
reevaluated with \nTwoPTwo.  The comparisons cover GPU direct and moment
evaluation at all three sizes and CPU evaluation at 5184 atoms.
\Cref{tab:gpu-validation} reports the
largest differences over these comparisons.  The energy difference per atom
is \(\Delta E/N\), and the force difference follows
\cref{eq:validation-errors}.  At 24,000 atoms, these differences remain
below \(3.2\times10^{-11}\)~eV/atom and
\(1.5\times10^{-11}\)~eV/\AA, respectively.
The acceptance limits specified before the comparisons were
\(10^{-8}\)~eV/atom and \(10^{-8}\)~eV/\AA.

\begin{table}[t]
  \centering
  \small
  \caption{Maximum differences between the TiO$_2$ production calculations and
  \nTwoPTwo\ evaluated at identical coordinates.  Maxima are taken over the
  direct and moment paths and the tested configurations, including CPU
  calculations at 5184 atoms and GPU calculations at all three sizes.}
  \label{tab:gpu-validation}
  \begin{tabular}{@{}rrr@{}}
    \toprule
    Atoms & \(\Delta E/N\) (eV/atom) & \(\Delta F\) (eV/\AA) \\
    \midrule
    192 & \(9.09\times10^{-13}\) & \(3.56\times10^{-13}\) \\
    5184 & \(6.83\times10^{-12}\) & \(4.78\times10^{-12}\) \\
    24,000 & \(3.18\times10^{-11}\) & \(1.50\times10^{-11}\) \\
    \bottomrule
  \end{tabular}
\end{table}

The MoS$_2$-confined water model was also checked against \nTwoPTwo\ at
the initial configuration of each system size.  A separate 1~ps NVE test
with a 0.5~fs time step gave maximum energy excursions of
\(6.57\times10^{-5}\) and \(6.61\times10^{-5}\)~eV/atom for LAMMPS with
\AccelNet\ and CabanaMD, respectively.

\section{AccelNet software and benchmark protocol}
\label{sec:implementation}

\AccelNet\ is an object-oriented software package written in Fortran 2008
for evaluating neural-network interatomic potentials.
GPU acceleration is implemented using OpenMP target directives, allowing
the CPU and GPU implementations to share the same numerical kernels.  It
provides a common descriptor and neural-network evaluator for supported
\aenet\ and \nTwoPTwo\ models, object-based Fortran interfaces, an
\aenet-style Fortran/C atomic-environment API, model-conversion utilities, and
a LAMMPS pair style \cite{plimpton1995lammps}.  It is an inference package:
training, reference-data preparation, and optimization remain the
responsibility of \aenet, \aenet-PyTorch, or \nTwoPTwo.  The source code is
publicly available at \url{https://github.com/cometscome/AccelNet}.

\subsection{Model input and compatibility}

For \aenet, \AccelNet\ reads the ASCII neural-network representation and
compatible native Fortran sequential-unformatted binary files.  Descriptor
metadata embedded in a supported network can reconstruct Chebyshev or
Behler2011 G1--G5 inputs, and all \aenet\ 2.0.4 activation codes 0--4 are
implemented.  Because the binary format depends on compiler record markers,
endianness, and kind representation, ASCII files provide the portable exchange
format.

A supported \nTwoPTwo\ short-range 2G model directory can be loaded directly,
without conversion.  The reader uses \texttt{input.nn}, the conventional
\texttt{weights.\%03d.data} files, and \texttt{scaling.data} when scaling is
enabled.  It supports symmetry-function types 2, 3, 9, 12, 13, and 20--25,
cutoff types 0--8,
and the fractional cutoff introduced by Mori et al.\ \cite{mori2023dynamic},
represented internally as cutoff type 9.
All activation functions in \nTwoPTwo\ 2.3.0, the documented scaling modes,
energy normalization, atomic reference energies, global or element-specific
network topologies, and \texttt{normalize\_nodes} are supported.  Node normalization is
folded exactly into the corresponding weights and biases during loading.
This includes weighted and compact symmetry functions.  Fourth-generation
and charge models and charge equilibration are not supported.  Unsupported settings are
rejected explicitly rather than evaluated with altered semantics.

The supplied Fortran converter maps the common supported subset between
\nTwoPTwo\ directories and the extended \AccelNet\ ASCII representation.  It
also reorders scaling arrays and first-layer weights when the descriptor
ordering conventions differ.  Model conversion is not required for direct
\nTwoPTwo\ directory loading.

\subsection{Evaluation interfaces and algorithm selection}

The structure-level predictor evaluates energies and analytical Cartesian
forces for in-memory structures and for XSF input; molecular or periodic
structures can also be read from \nTwoPTwo\ \texttt{input.data} files.  The
atomic API accepts a central atom and its local neighbors, allowing existing
codes organized around the \aenet\ calling convention to replace the
inference library without changing their surrounding atom loop.

For Chebyshev angular descriptors, the public API exposes \texttt{direct},
\texttt{moment}, and \texttt{auto} evaluation modes.  In version 1.1.0,
automatic selection estimates the work from the angular-neighbor count,
polynomial order, and number of moments.  For the applicable \nTwoPTwo\
angular functions, automatic selection uses moments for exact integer
powers \(1\le\zeta\le10\) with at least 16 angular neighbors.  Explicit
moment selection also permits powers through 16 and removes the
neighbor-count threshold in the shared CPU/GPU evaluator.  The retained
atomic reference evaluator uses moments only through order ten.
Other terms use direct evaluation, so one model
may use both paths.
Functions sharing the same cutoff radius and \((\eta,r_s)\) radial factor are
grouped so that their species-resolved moments can be reused across
\(\lambda\), \(\zeta\), and neural-network input channels.  Algorithm
selection changes neither descriptor scaling nor network parameters.

\subsection{Build and library distribution}

The standard build requires CMake 3.20 or newer, a Fortran 2008-compatible
compiler, and a C compiler.  The core libraries do not require BLAS, LAPACK,
MPI, \aenet, or \nTwoPTwo; those packages are used only by selected reference
tests or external integrations.  The build produces separate descriptor and
predictor libraries, command-line programs for prediction and model
conversion, a C header, Fortran module files, and CMake package metadata for
downstream projects.  The same CMake configuration builds the unit and
integration tests executed through CTest.  GPU support is enabled separately
and uses a Fortran compiler with OpenMP target support.  The NVIDIA
implementation used here was built with NVIDIA HPC SDK 25.3 and CUDA 12.8.

\subsection{LAMMPS integration}

Source interfaces are supplied for LAMMPS 4Feb2020 through its traditional
make package and for LAMMPS 29Aug2024 Update 4 and 22Jul2025 Update 6 through
CMake packages.  The CPU interfaces provide \texttt{pair\_style accelnet}
and call the same inference library.  An
\aenet/\AccelNet\ model is specified by its element network files.  A
\nTwoPTwo\ model is specified by its directory together with an explicit map
from LAMMPS atom types to element names; it is loaded without an intermediate
model file.  Every MPI rank reads the same model input, while neighbor data are
provided by LAMMPS.  Model loading and file parsing are completed before
the timed MD region.

\subsection{GPU implementation}
\label{sec:gpu-implementation}

The GPU evaluator processes a batch of central atoms using Fortran OpenMP
target directives.  Each atom's neighbors occupy a segment of a shared
index array, with offsets marking the segment boundaries.  Bond
displacements are stored alongside these indices.  Descriptor evaluation,
neural-network evaluation, and force and virial accumulation run on the
GPU in double precision.  Model parameters and work arrays remain on the
device between calls.

In the moment path, functions with common radial parameters share
species-resolved moments and their contractions.  After neural-network
backpropagation, the energy derivatives with respect to the descriptors
are combined with these moments to form coefficients of a polynomial in
the unit bond direction,
\begin{equation}
  \mathcal{P}(u_x,u_y,u_z)
  =\sum_{\substack{a,b,c\ge0\\a+b+c\le P}} A_{abc}u_x^a u_y^b u_z^c.
  \label{eq:force-polynomial}
\end{equation}
The indices \(a,b,c\) are integers, \(P\) is the maximum angular degree, and the coefficients
\(A_{abc}\) are prepared for each central atom, radial group, and neighbor
species.  The polynomial and its three derivatives are evaluated together
by nested Horner recurrences.  For one variable \(t\), an intermediate
value \(v\) and its derivative \(d\) are updated as
\begin{equation}
  v_{\mathrm{new}}=t v_{\mathrm{old}}+b,\qquad
  d_{\mathrm{new}}=t d_{\mathrm{old}}+v_{\mathrm{old}},
  \label{eq:horner-derivative}
\end{equation}
where \(b\) is the next polynomial coefficient.  Applying this recurrence
successively along the three Cartesian components supplies the angular
part of the force.  The radial derivatives and self-pair corrections are
then combined with it in the neighbor loop.
\ref{app:moment-forces} derives the coefficients and the complete bond
derivative, including mixed-species contractions and force assembly.

The LAMMPS interface, \texttt{pair\_style accelnet/gpu}, couples these
kernels to the LAMMPS GPU package.  With GPU neighbor construction, the
adapter uses the package's device coordinates and neighbor arrays to form
the batched representation.  Forces are returned to LAMMPS after evaluation
for integration and interprocess communication.  The interface reads
network files containing descriptor metadata; \nTwoPTwo\ models are
prepared with the supplied converter, which preserves the descriptor
definitions, scaling, and network parameters.

\subsection{Benchmark systems and timing protocol}

The single-core measurements use the first three models in
\cref{tab:models}.
\cref{tab:neighbors} reports the angular-neighbor distribution in each
initial configuration.  The anatase and rutile configurations have
different neighbor distributions despite their common
angular cutoff of 6.5~\AA.  All periodic images inside the angular
cutoff are counted.  The pair-count column is the atom average of
\(z_i(z_i-1)/2\), not the value formed from the mean \(z_i\).

\begin{table}[t]
  \centering
  \small
  \setlength{\tabcolsep}{3.5pt}
  \caption{Angular-neighbor statistics computed from the initial benchmark
  configurations, including periodic images.  \(P_{90}\) is the 90th percentile.}
  \label{tab:neighbors}
  \begin{tabularx}{\linewidth}{@{}Xrrrrrr@{}}
    \toprule
    System & \(N\) & \(r_c\) (\AA) & Mean \(z\) & Med. \(z\) & \(P_{90}(z)\) & Mean pairs \\
    \midrule
    H$_2$O Chebyshev & 1152 & 3.175 & 11.49 & 11 & 14 & 62.68 \\
    TiO$_2$ Chebyshev\newline (anatase) & 1152 & 6.500 & 92.00 & 93 & 93 & 4187.00 \\[3pt]
    TiO$_2$ \nTwoPTwo\ model\newline (rutile) & 192 & 6.500 & 100.25 & 100 & 102 & 4976.25 \\
    \bottomrule
  \end{tabularx}
\end{table}

The single-core results in \cref{sec:cpu-results} use \AccelNet\ 1.1.0 on
an Intel Xeon Gold 6526Y.  Each process was pinned to one physical core.
All implementations were called through the same LAMMPS executable,
using \aenet\ 2.0.4 and \nTwoPTwo\ as external references.

The time step is 0.5~fs for H$_2$O
and 1~fs for both TiO$_2$ models.  H$_2$O and TiO$_2$ Chebyshev use
stochastic velocity-rescaling temperature control at 300~K; the
\nTwoPTwo\ TiO$_2$ model uses NVE integration from a 300~K initialization.
The CPU direct and moment paths use the same model and executable; only
the angular evaluation algorithm changes.

The GPU measurements in \cref{sec:gpu-results} use \AccelNet\ 1.1.0,
LAMMPS 22Jul2025 Update 6, and one NVIDIA H100 NVL with double-precision
arithmetic.  TiO$_2$ trajectories start at
300~K and use NVE integration with a 1~fs time step.
Direct and moment evaluation use the same model and executable.
At 5184 atoms, the same TiO$_2$ workload is also evaluated with \nTwoPTwo\
and both \AccelNet\ CPU paths on one Xeon Gold 6526Y core.  All five
methods use identical initial coordinates and velocities, integration
settings, and output requests.

For MoS$_2$-confined water, the time step is 0.5~fs.  The GPU
comparison uses CabanaMD's \texttt{TEAM\_VECTOR} setting
\cite{desai2022cabanamd}.  Both codes use NVE integration from the same
initial positions and velocities.  The 495-atom CPU comparison uses one
Intel Xeon Gold 6526Y core.

Each run comprises 100 warm-up steps followed by 500 measured steps for
the H$_2$O and TiO$_2$ benchmarks, or 200 measured steps for
MoS$_2$-confined water.  Reported timings are medians over five independent
runs for H$_2$O and TiO$_2$, and three for MoS$_2$-confined water.
All timings exclude initialization and warm-up.
Computational details and integration parameters are given in
\ref{app:benchmark-details}.

\section{Results}
\label{sec:results}

\subsection{Single-core LAMMPS wall times}
\label{sec:cpu-results}

\begin{figure*}[t]
  \centering
  \includegraphics[width=\textwidth]{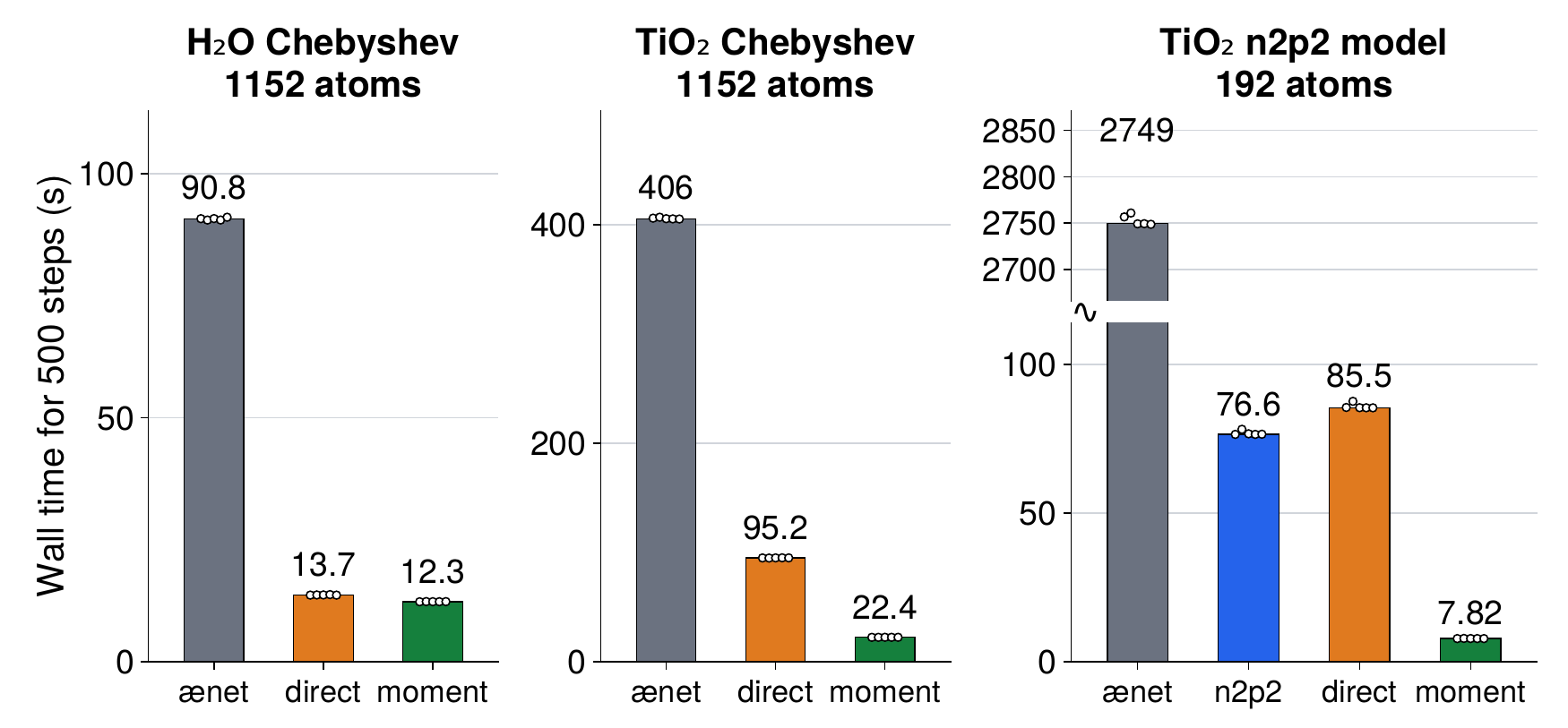}
  \caption{Wall time for the 500-step measured region of LAMMPS execution.
  All cases include atomic motion.
  Bars show the five-trial median and open circles show individual process
  launches.  The vertical axis of the 192-atom TiO$_2$ panel is broken so that
  the \aenet\ result can be shown together with the substantially faster
  \nTwoPTwo\ and \AccelNet\ results.  Each measurement used one pinned
  Xeon Gold 6526Y core and a GCC/gfortran 11.4 build.}
  \label{fig:wall-times}
\end{figure*}

The complete trial distributions and medians are shown in
\cref{fig:wall-times}; numerical values are listed in \cref{tab:performance}.
Each median is calculated from five independent process launches.

\begin{table*}[t]
  \centering
  \small
  \caption{Single-core LAMMPS timings for the protocols in
  \cref{fig:wall-times}.  \(S_\mathrm{eng}\) is
  external/direct, \(S_\mathrm{mom}\) is direct/moment, and
  \(S_\mathrm{tot}\) is external/moment.  For the 192-atom TiO$_2$ model,
  \nTwoPTwo\ is used as the external reference; the \aenet\ row additionally
  reports its ratio to the same \AccelNet\ moment timing.}
  \label{tab:performance}
  \begin{tabular}{@{}llrrrr@{}}
    \toprule
    System & Implementation & Median (s) & \(S_\mathrm{eng}\) & \(S_\mathrm{mom}\) & \(S_\mathrm{tot}\) \\
    \midrule
    H$_2$O Chebyshev & \aenet & 90.7700 & -- & -- & -- \\
                     & \AccelNet\ direct & 13.6924 & 6.629 & -- & -- \\
                     & \AccelNet\ moment & 12.3113 & -- & 1.112 & 7.373 \\
    \midrule
    TiO$_2$ Chebyshev & \aenet & 405.752 & -- & -- & -- \\
                      & \AccelNet\ direct & 95.1566 & 4.264 & -- & -- \\
                      & \AccelNet\ moment & 22.4181 & -- & 4.245 & 18.099 \\
    \midrule
    TiO$_2$ \nTwoPTwo\ model & \aenet & 2749.48 & -- & -- & 351.657 \\
                               & \nTwoPTwo & 76.5967 & -- & -- & -- \\
                               & \AccelNet\ direct & 85.5245 & 0.896 & -- & -- \\
                               & \AccelNet\ moment & 7.81865 & -- & 10.939 & 9.797 \\
    \bottomrule
  \end{tabular}
\end{table*}

\subsection{Separating engine and algorithmic gains}

Let \(T_{\mathrm{ext}}\), \(T_{\mathrm{direct}}\), and
\(T_{\mathrm{moment}}\) denote the external-code, optimized-direct, and
moment wall times.  The end-to-end speedup factors exactly as
\begin{equation}
  \frac{T_{\mathrm{ext}}}{T_{\mathrm{moment}}}
  = \frac{T_{\mathrm{ext}}}{T_{\mathrm{direct}}}
    \frac{T_{\mathrm{direct}}}{T_{\mathrm{moment}}}.
  \label{eq:speedup-decomposition}
\end{equation}
The second factor isolates the effect of moment factorization within
\AccelNet.

For H$_2$O, the direct-to-moment ratio is 1.112, and the end-to-end
speedup over \aenet\ is 7.373.  Most of this improvement is already
present in the direct evaluator.

For TiO$_2$ Chebyshev, moment evaluation is 4.245 times faster than direct
evaluation.  The end-to-end factor of 18.099 relative to \aenet\ decomposes into
\begin{equation}
  18.099 \simeq 4.264\times 4.245.
\end{equation}

The TiO$_2$ \nTwoPTwo\ model gives the largest CPU direct-to-moment ratio,
10.939.  Direct \AccelNet\ evaluation takes 85.525~s, compared with
76.597~s for \nTwoPTwo, while the moment path takes 7.819~s.  The resulting
speedup over \nTwoPTwo\ is 9.797.  Evaluation of the same model through
\aenet\ takes 2749.48~s.  The decomposition uses \nTwoPTwo\ as the
reference because the model was trained with that implementation.

\FloatBarrier
\subsection{GPU performance}
\label{sec:gpu-results}

\subsubsection{Moment evaluation for \texorpdfstring{TiO$_2$}{TiO2}}

\begin{figure}[!htbp]
  \centering
  \includegraphics[width=0.90\linewidth]{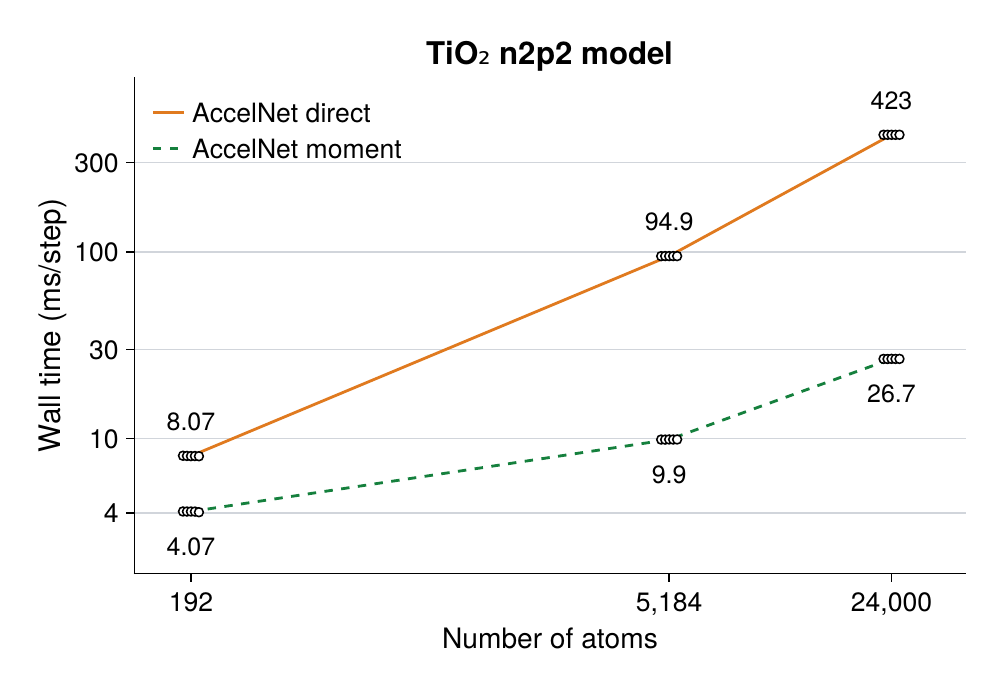}
  \caption{GPU direct and moment evaluation of the TiO$_2$ \nTwoPTwo\ model
  in LAMMPS with \AccelNet\ 1.1.0.  Lines connect the median wall-clock
  times per MD step on one H100 NVL.  Open circles show five independent launches;
  error bars give their minimum and maximum.  Both axes are logarithmic.}
  \label{fig:gpu-tio2}
\end{figure}

\Cref{fig:gpu-tio2} compares direct and moment evaluation on the H100 NVL
for the \nTwoPTwo\ TiO$_2$ model.  At 192 atoms, the median wall-clock times
are 8.074 and 4.074~ms/step, respectively.  The difference grows at larger
system sizes: 94.860 versus 9.895~ms/step at 5184 atoms, and 422.878 versus
26.713~ms/step at 24,000 atoms.  The corresponding direct-to-moment ratios
are 1.98, 9.59, and 15.83.

\Cref{tab:tio2-cpu-gpu} compares all five evaluation routes for the
5184-atom system.  The CPU direct-to-moment ratio is 10.74, while the
corresponding GPU ratio is 9.59.  Moving the moment evaluator from one
CPU core to the GPU gives a further factor of 43.45.  The total ratio
relative to \nTwoPTwo\ on one CPU core is 431.22, including the factor
of 0.924 between \nTwoPTwo\ and \AccelNet\ CPU direct evaluation.
The CPU-to-GPU comparison includes the change from GCC/gfortran to the
NVIDIA compiler and device backend.

\begin{table}[!htbp]
  \centering
  \small
  \caption{Wall-clock time per MD step for the 5184-atom TiO$_2$ system.
  Each row contains five independent launches with identical initial
  conditions and integration settings.  CPU rows use one Xeon Gold
  6526Y core; GPU rows use one H100 NVL.}
  \label{tab:tio2-cpu-gpu}
  \begin{tabular}{@{}lrr@{}}
    \toprule
    Implementation & Median (ms/step) & Min--max (ms/step) \\
    \midrule
    \nTwoPTwo\ CPU & 4267.160 & 4263.820--4271.020 \\
    \AccelNet\ CPU direct & 4616.960 & 4612.640--4621.380 \\
    \AccelNet\ CPU moment & 429.986 & 428.758--430.328 \\
    \AccelNet\ GPU direct & 94.860 & 94.796--94.888 \\
    \AccelNet\ GPU moment & 9.895 & 9.885--9.904 \\
    \bottomrule
  \end{tabular}
\end{table}

The low angular degrees and shared radial parameters make this model
well suited to moment evaluation.  Degrees through four require 35
Cartesian monomials.  Their moments are reused across the 54 angular
functions per element, and the force contraction described in
\cref{sec:gpu-implementation} retains a single-neighbor traversal.
The measured reduction in wall-clock time shows that eliminating pair work
remains effective when both evaluation paths run on the GPU.
The replicas have the same density and cutoff, so the local neighbor
count remains approximately constant across these system sizes.

\FloatBarrier
\subsubsection{\texorpdfstring{MoS$_2$}{MoS2}-confined water}

\begin{figure}[!htbp]
  \centering
  \includegraphics[width=\textwidth]{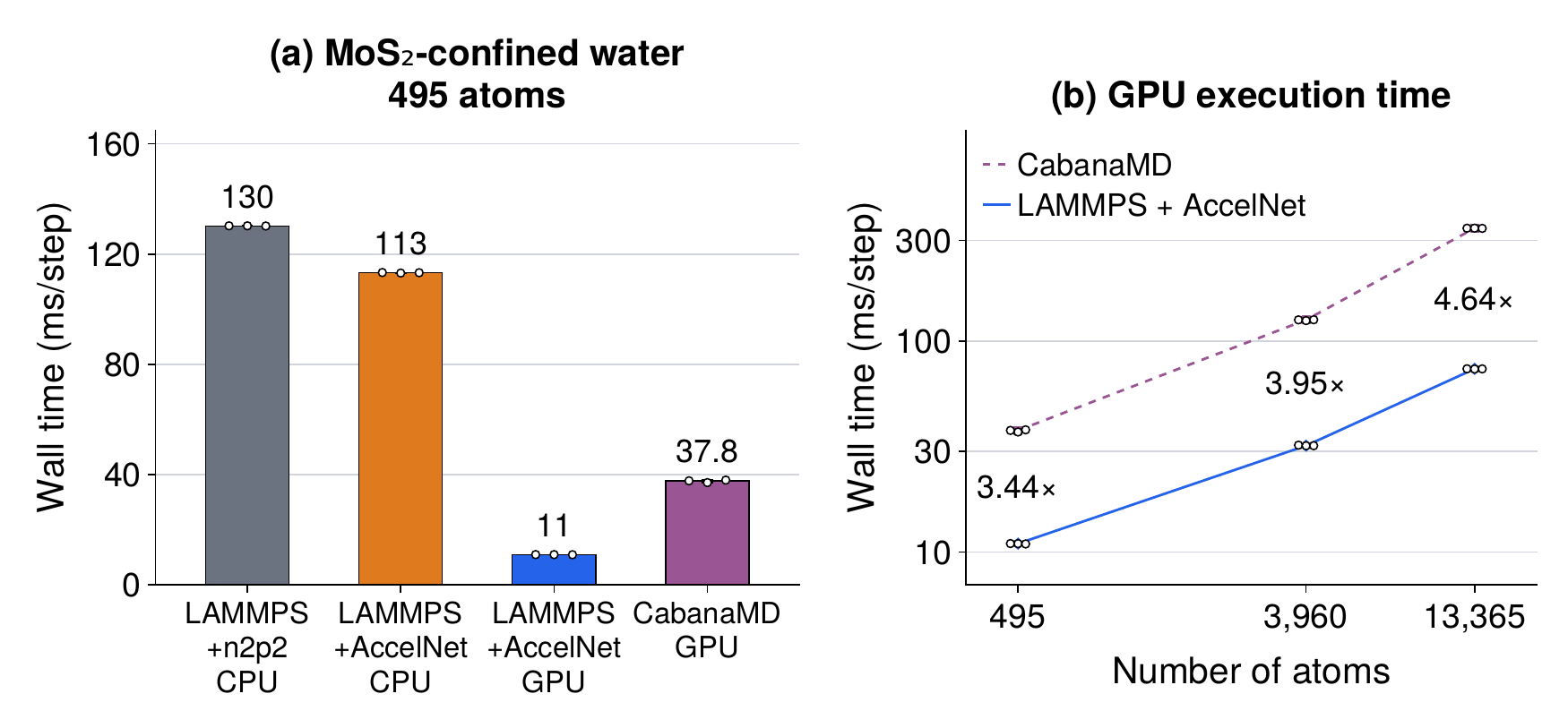}
  \caption{Wall-clock time per MD step for MoS$_2$-confined water using the published
  potential of Schran et al.\ \cite{schran2021aqueous}.
  (a) CPU and GPU measurements at 495 atoms.
  (b) GPU measurements for LAMMPS with \AccelNet\ (direct) and CabanaMD
  (\texttt{TEAM\_VECTOR}); annotations give the CabanaMD-to-\AccelNet\
  time ratio.  The CPU is one Xeon Gold 6526Y core and the GPU is one
  H100 NVL.  Bars and lines show medians of three independent launches,
  open circles show individual times, and error bars give their range.
  Panel (b) uses logarithmic axes.}
  \label{fig:gpu-mos2}
\end{figure}

The published MoS$_2$-confined water model provides a four-element test of
GPU direct evaluation.  \Cref{fig:gpu-mos2} compares LAMMPS coupled to
\AccelNet\ with CabanaMD, which implements neural-network potentials
using the Kokkos and Cabana libraries \cite{desai2022cabanamd}.
At 495 atoms, LAMMPS with \AccelNet\ takes 10.974~ms/step on the H100 NVL,
compared with 113.263~ms/step on one CPU core and 37.750~ms/step for
CabanaMD on the same GPU.  At 3960 and 13,365 atoms, the GPU times are
31.961 and 73.885~ms/step for LAMMPS with \AccelNet, and 126.100 and
342.900~ms/step for CabanaMD.  Thus LAMMPS with \AccelNet\ is faster by
factors of 3.44--4.64 in this comparison.  These MD timings
demonstrate the performance of the GPU interface for an existing
multi-element potential whose angular functions use direct pair evaluation.

\FloatBarrier
\section{Discussion}
\label{sec:discussion}

\subsection{Evaluation of existing trained models}

The model comparisons in \cref{sec:verification} support using the tested
H$_2$O and TiO$_2$ potentials through \AccelNet\ without retraining.
The descriptor-level tests isolate errors in the moment evaluation, while
the external-model comparisons check the energy and force predictions
after input scaling and network evaluation.

\subsection{Interpretation of the measured speedups}

The measurements follow the cost balance derived in
\cref{sec:computational-cost}.  The short-cutoff H$_2$O model offers little
pair work to remove, whereas the longer-cutoff TiO$_2$ environments favor
moment evaluation.  The cost also depends on angular degree and on the
number of radial and species groups.  Sharing moments across descriptors
reduces both their construction cost and the subsequent force contraction.

The GPU results show that this reduction in work remains useful under
parallel execution.  Moment construction and force contraction distribute
work across central atoms and moment components.  The low-degree TiO$_2$
model combines this parallelism with substantial reuse of the moments.
For the MoS$_2$-confined water model, the benefit instead comes from the
GPU implementation of direct evaluation and its integration into LAMMPS.

Comparing direct and moment paths within \AccelNet\ separates the effect
of the factorization from differences between inference engines.  The
external-code comparisons additionally include data layout, network
evaluation, neighbor handling, and the surrounding MD implementation.

\subsection{Applicability and limitations}

Exact factorization requires the two-neighbor dependence to enter through a
finite polynomial of \(\cos\theta_{ijk}\), with the remaining neighbor
weights separable between \(j\) and \(k\).  The Chebyshev angular descriptor
and the supported positive-integer-power \nTwoPTwo\ functions satisfy these
conditions.  Angular functions with a nonseparable dependence on \(r_{jk}\),
noninteger angular powers, or general nonpolynomial kernels do not.  \AccelNet\
retains direct evaluation for such terms, thereby preserving the model rather
than replacing the unsupported function by an approximation.

The linear dependence on neighbor count is obtained at fixed polynomial order
and fixed channel structure.  The number of Cartesian moments through degree
\(P\) is \(\binom{P+3}{3}\), so initialization, contraction, storage, and
force differentiation become increasingly expensive at high degree.  Other
finite bases, including spherical or symmetric-traceless tensor forms, may
reduce Cartesian redundancy, but their practical value would depend on the
cost of basis transformations and on maintaining exact compatibility with
the stored descriptors.

\section{Conclusions}
\label{sec:conclusions}

We have derived an exact moment representation for angular descriptors
with separable one-neighbor weights and finite polynomial dependence on
the cosine of the bond angle.  With
\(\cos\theta_{ijk}=\hat{\vect r}_{ij}\mathbin{\cdot}\hat{\vect r}_{ik}\),
each dot-product power has a finite Cartesian feature representation.  The
pair sum can therefore be reconstructed from one-neighbor moments together
with an exact subtraction of the self contributions.  This reformulation
changes the evaluation algorithm but not the descriptor itself.

We applied the reformulation to the Artrith--Urban--Ceder Chebyshev angular
descriptor used by \aenet\ and to the supported polynomial angular functions
used by \nTwoPTwo.  The resulting software, \AccelNet, reads existing trained
models, preserves their descriptor ordering and scaling, network parameters,
energies, and analytical forces, and provides a LAMMPS interface.  Tests for
H$_2$O and TiO$_2$ reproduce the reference energies and forces to
floating-point roundoff without regenerating training data or retraining the
neural networks.

The CPU direct-to-moment ratios are 1.11 for H$_2$O, 4.24 for TiO$_2$
Chebyshev, and 10.94 for the TiO$_2$ \nTwoPTwo\ model.  On an H100 NVL GPU, moment
evaluation reduces the wall-clock time per MD step of the 24,000-atom TiO$_2$
\nTwoPTwo\ model from 422.9 to 26.7~ms/step, a factor of 15.8.  The
MoS$_2$-confined water benchmark further demonstrates GPU execution of a
published four-element potential through LAMMPS.  Together, these results
establish a CPU/GPU evaluation route that combines existing trained
potentials with the reduced cost of exact moment factorization wherever
their angular functions admit it.

\section*{Software availability}

The \AccelNet\ source code, model-conversion tools, and LAMMPS interfaces are
available at \url{https://github.com/cometscome/AccelNet}.  The GPU extension
is included in version 1.1.0.
Original \AccelNet\
code is distributed under the MIT License.  The retained \aenet-derived
linked-cell source and the LAMMPS pair-style sources preserve their MPL-2.0
and GPL-2.0 licenses, respectively, with full notices included in the
repository.

\section*{CRediT authorship contribution statement}

\textbf{Yuki Nagai:} Conceptualization, Methodology, Software, Validation,
Investigation, Data curation, Visualization, Writing -- original draft,
Writing -- review and editing.

\section*{Declaration of competing interest}

The author declares no competing financial interest.

\section*{Acknowledgements}

The work of Y.N. was partially supported by JSPS KAKENHI Grant Numbers
22H05114, 26K00651, and 26K01464.  This work was partially supported by
``Joint Usage/Research Center for Interdisciplinary Large-scale Information
Infrastructures (JHPCN)'' in Japan (Project ID: jh250046).

\input{force-appendix}
\input{benchmark-appendix}

\bibliographystyle{elsarticle-num}
\setlength{\bibsep}{2pt}
\bibliography{references}

\end{document}

%% file: verification-details.tex
We quantified the descriptor and derivative errors using the retained atomic
evaluators in release 1.1.0, compiled in double precision with GNU Fortran
15.1.0 and \texttt{-O3} on an ARM64 CPU.  A deterministic environment contains
20 neighbors of two alternating species at radii 1.00--1.88~\AA.  Ten cutoff
functions are tested.  The Chebyshev tests use radial order four and angular
order six with descriptor versions 0, 1, and 10, giving 30 configurations of
the descriptor definition on this geometry.  The \nTwoPTwo{}-form tests use
four wide angular functions with \(\zeta=1,2,3,10\), same- and mixed-species
pairs, and radial shifts of 0.20, 0.55, 0.80, and 1.10~\AA.

\Cref{tab:derivative-validation} reports maxima for unscaled descriptors
and their derivatives.  Each derivative is recovered from the moment
contraction by setting one descriptor coefficient to unity and all others
to zero, then compared with the corresponding explicit-pair derivative.
Both central and neighbor derivatives are included.  For an independent
energy--force check, we use the scalar test network
\(E=(1~\mathrm{eV})\tanh[0.1+\sum_\mu w_\mu G_\mu]\), with
\(w_\mu=0.01(-1)^\mu\).  All 60 neighbor-coordinate force components are
compared with the centered difference
\begin{equation}
 F_q^{\mathrm{FD}}=-\frac{E(\vect r+h\vect e_q)-E(\vect r-h\vect e_q)}{2h},
 \qquad h=10^{-6}~\text{\AA},
 \label{eq:force-finite-difference}
\end{equation}
where \(\vect e_q\) selects the displaced coordinate.  Both displaced
energies use moment evaluation.

\begin{table}[!htbp]
 \centering\small
 \caption{Measured maximum absolute errors in the synthetic atomic-evaluator
 tests.  \(\Delta G\) compares unscaled descriptors, \(\Delta J\) compares
 every Cartesian descriptor derivative (\AA$^{-1}$), and
 \(\Delta F_{\mathrm{FD}}\) compares the test-network force with the centered
 energy difference (eV/\AA).  Separate acceptance limits were
 \(2\times10^{-11}\), \(3\times10^{-10}\)~\AA$^{-1}$, and
 \(2\times10^{-7}\)~eV/\AA, respectively.}
 \label{tab:derivative-validation}
 \begin{tabular}{@{}lrrr@{}}
 \toprule
 Family & \(\max|\Delta G|\) & \(\max|\Delta J|\) & \(\max|\Delta F_{\mathrm{FD}}|\)\\
 \midrule
 Chebyshev & \(9.09\times10^{-12}\) & \(2.27\times10^{-12}\) & \(2.87\times10^{-9}\)\\
 Wide angular & \(4.26\times10^{-14}\) & \(1.18\times10^{-14}\) & \(9.26\times10^{-9}\)\\
 \bottomrule
 \end{tabular}
\end{table}

Further tests of the wide angular functions cover every integer order
from 11 through 16 using the shared CPU/GPU evaluator.  The tests confirm
that the moment path is selected and include independent \nTwoPTwo\
comparisons for two 16-atom synthetic systems with two and four species.
Across ten model/mode comparisons on the H100, the maximum total
energy and force-component differences from \nTwoPTwo\ are
\(1.21\times10^{-14}\)~eV and
\(7.87\times10^{-19}\)~eV/\AA.  Centered energy differences from
\nTwoPTwo, using \(h\simeq1.05835\times10^{-5}\)~\AA, check the three force components
of the first atom in each system against forced-moment forces; the maximum
difference is \(3.01\times10^{-10}\)~eV/\AA.  The acceptance criterion
is \(|\Delta F|\le\varepsilon_{\mathrm{abs}}
+2\times10^{-6}|F^{\mathrm{FD}}|\), with
\(\varepsilon_{\mathrm{abs}}\simeq1.03\times10^{-6}\)~eV/\AA.
These finite sets of tests characterize the reported degrees and environments;
the algebraic identity alone does not bound cancellation errors for arbitrary
high-degree models.

%% file: force-appendix.tex
\appendix
\section{From moment contractions to atomic forces}
\label{app:moment-forces}

Consider the contribution of a fixed central atom \(i\), and suppress its
index below.  Within one radial group, let \(s\) label a chemical channel
and define
\begin{equation}
 m_{\boldsymbol\ell}(\vect u)=u_x^{\ell_x}u_y^{\ell_y}u_z^{\ell_z},\qquad
 M_{\boldsymbol\ell}^{s}=\sum_j h_j^s m_{\boldsymbol\ell}(\vect u_j),\qquad
 D^{st}=\sum_j h_j^s h_j^t.
\end{equation}
The weight \(h_j^s=h^s(r_{ij},\alpha_j)\) includes the cutoff and chemical
weight or species mask.  A descriptor \(\mu\) using channels \(a_\mu,b_\mu\)
has the form
\begin{equation}
 G_\mu=\kappa_\mu\left[
 \sum_{\boldsymbol\ell} c_{\mu\boldsymbol\ell}
 M_{\boldsymbol\ell}^{a_\mu}M_{\boldsymbol\ell}^{b_\mu}
 -P_\mu(1)D^{a_\mu b_\mu}\right],
 \qquad
 c_{\mu\boldsymbol\ell}
 =a_{\mu,|\boldsymbol\ell|}
 \frac{|\boldsymbol\ell|!}{\ell_x!\ell_y!\ell_z!},
 \label{eq:app-general-contraction}
\end{equation}
where \(P_\mu(c)=\sum_p a_{\mu p}c^p\).  Chebyshev descriptors have
\(a_\mu=b_\mu=\gamma\), \(\kappa_\mu=1/2\), and \(P_\mu=T_n\).
For the wide angular functions, \(a_\mu=\beta\), \(b_\mu=\delta\),
\(P_\mu(c)=(1+\lambda c)^\zeta\), and
\(\kappa_\mu=2^{1-\zeta}/(1+\delta_{\beta\delta})\).

\subsection{Coefficients supplied by network differentiation}

Write \(q_\mu=\partial E_i/\partial G_\mu\) for the derivative with
respect to the unscaled descriptor.  For affine input scaling
\(\widetilde G_\mu=s_\mu G_\mu+t_\mu\), backpropagation gives
\(q_\mu=s_\mu\partial E_i/\partial\widetilde G_\mu\), including any
output-energy conversion in \(E_i\).  Differentiating
\cref{eq:app-general-contraction} gives
\begin{align}
 A_{s\boldsymbol\ell}
 &=\sum_\mu q_\mu\kappa_\mu c_{\mu\boldsymbol\ell}
   \left[\delta_{s a_\mu}M_{\boldsymbol\ell}^{b_\mu}
        +\delta_{s b_\mu}M_{\boldsymbol\ell}^{a_\mu}\right],
 \label{eq:app-moment-adjoint}\\
 \dd E_i\big|_{\rm ang}
 &=\sum_{s,\boldsymbol\ell}A_{s\boldsymbol\ell}\dd M_{\boldsymbol\ell}^s
 -\sum_\mu q_\mu\kappa_\mu P_\mu(1)\dd D^{a_\mu b_\mu}.
 \label{eq:app-energy-differential}
\end{align}
The deltas in \cref{eq:app-moment-adjoint} are Kronecker symbols.  When
the two channels coincide, both contributions enter the same coefficient,
giving \(2q_\mu\kappa_\mu c_{\mu\boldsymbol\ell}M_{\boldsymbol\ell}^{a_\mu}\).
For distinct species, channel \(\beta\) receives the moment of \(\delta\)
and vice versa; their disjoint masks make \(D^{\beta\delta}=0\).
Thus the same formula includes the same-species factor of two and the
mixed-species product without changing the pair normalization.

Define the force polynomial for channel \(s\) by
\begin{equation}
 Q_s(\vect u)=\sum_{\boldsymbol\ell}A_{s\boldsymbol\ell}
 m_{\boldsymbol\ell}(\vect u).
 \label{eq:app-force-polynomial}
\end{equation}
These are the coefficients denoted \(A_{abc}\) in
\cref{eq:force-polynomial}, for the specified group and channel.  They are
evaluated once at the current configuration and then used as coefficients
in the first differential \cref{eq:app-energy-differential}.

\subsection{Bond derivatives and self-pair correction}

With \(\vect r_{ij}=\vect r_j-\vect r_i\),
\(r_{ij}=|\vect r_{ij}|\), and \(\vect u_j=\vect r_{ij}/r_{ij}\),
\begin{equation}
 \frac{\partial r_{ij}}{\partial\vect r_{ij}}=\vect u_j,\qquad
 \frac{\partial\vect u_j}{\partial\vect r_{ij}}
 =\frac{1}{r_{ij}}\left(I-\vect u_j\vect u_j^{\mathsf T}\right),
 \label{eq:app-direction-derivative}
\end{equation}
where \(I\) is the three-dimensional identity matrix.  Let
\(h_j^{s\prime}=\partial h^s(r_{ij},\alpha_j)/\partial r_{ij}\).
Substituting the moment differentials into
\cref{eq:app-energy-differential} yields the angular contribution
\begin{align}
 \vect g_{ij}^{\rm ang}
 ={}&\sum_s\left[
 h_j^{s\prime}Q_s(\vect u_j)\vect u_j
 +\frac{h_j^s}{r_{ij}}
 (I-\vect u_j\vect u_j^{\mathsf T})\nabla_{\vect u}Q_s(\vect u_j)
 \right]\nonumber\\
 &-\sum_\mu q_\mu\kappa_\mu P_\mu(1)
 \left(h_j^{a_\mu\prime}h_j^{b_\mu}
      +h_j^{a_\mu}h_j^{b_\mu\prime}\right)\vect u_j.
 \label{eq:app-bond-force}
\end{align}
The last line differentiates the self-pair subtraction.  This term has
no contribution from variations of the unit bond direction because
\(\vect u_j\cdot\vect u_j=1\); its radial derivative remains.
For a same-species channel the radial factor is \(2h_j^s h_j^{s\prime}\);
for disjoint species it is zero.  For Chebyshev weights,
\(h_j^{\gamma\prime}=w_{\alpha_j}^{(\gamma)}f_c'(r_{ij})\).
For the wide angular weights,
\begin{equation}
 h_j^{s\prime}=\chi_s(\alpha_j)e^{-\eta(r_{ij}-r_{\mathrm{s}})^2}
 \left[f_c'(r_{ij})-2\eta(r_{ij}-r_{\mathrm{s}})f_c(r_{ij})\right].
\end{equation}
In this last expression \(r_{\mathrm{s}}\) denotes the radial shift, as in
\cref{eq:n2p2-one-neighbor-weight}.

\subsection{Force assembly}

The nested Horner recurrence in \cref{eq:horner-derivative} evaluates
\(Q_s\) and \(\nabla_{\vect u}Q_s\) in
\cref{eq:app-bond-force}.  Contributions from all radial groups, ordinary
radial descriptors, and any directly evaluated angular terms are added to
obtain \(\vect g_{ij}=\partial E_i/\partial\vect r_{ij}\).  The force
increments from this central energy are
\begin{equation}
 \vect F_i\mathrel{+}=\vect g_{ij},\qquad
 \vect F_j\mathrel{-}=\vect g_{ij}.
 \label{eq:app-force-assembly}
\end{equation}
Summing these increments over central atoms differentiates
\(E=\sum_i E_i\).  After the moments and their coefficients have been
formed, \cref{eq:app-bond-force} requires one traversal of the neighbors
for each radial group.

%% file: benchmark-appendix.tex
\section{Computational conditions}
\label{app:benchmark-details}

\subsection{Computational environment}

The CPU calculations used GCC and GNU Fortran 11.4 with Release builds.
The \AccelNet\ C++ and Fortran release settings used \texttt{-O3};
OpenBLAS 0.3.21.dev used its pthread implementation with one thread.
The single-core LAMMPS calculations used no OpenMP parallelism.
For the TiO$_2$ GPU calculations, NVIDIA HPC SDK 25.3 and CUDA 12.8 were
used.  The Fortran evaluator was compiled with \texttt{-fast -O3} and
OpenMP target flags \texttt{-mp=gpu -gpu=cc90}.  The GPU-linked LAMMPS
C++ build used \texttt{-O2 -DNDEBUG}.  Calculations used one host thread
and double-precision arithmetic, with GPU offload required.

\subsection{Integration and timing}

For the single-core H$_2$O and TiO$_2$ Chebyshev benchmarks, velocities
were initialized at 300~K and temperature was controlled by stochastic
velocity rescaling.  The relaxation times were 1~ps and
0.001~ps, respectively.  Total linear momentum was removed every 100
steps.  The time steps were 0.5~fs for H$_2$O and 1~fs for TiO$_2$.
The \nTwoPTwo\ TiO$_2$ benchmarks instead used NVE integration,
initialized at 300~K, with a 1~fs time step.  Within each comparison,
all implementations used the same initial coordinates and velocities.

The TiO$_2$ CPU/GPU comparison used a 0.6~\AA\ neighbor skin, rebuild
checks every ten steps, and atom sorting every 100 steps.  Thermodynamic
quantities and the configurational virial were requested every 100
steps.  Initialization, warm-up, validation, and per-atom file output
were outside the timed region.

The MoS$_2$-confined water comparison used a 0.6~\AA\ neighbor skin,
rebuilding every ten steps, and thermodynamic output every 100 steps.
For both codes, the measured interval comprised steps 101--300, following
100 warm-up steps.